\documentclass[aps,prb,twocolumn]{revtex4-1}
\usepackage{graphicx}

\begin{document}

\title{Pulling Platinum Atomic Chains by Carbon Monoxide Molecules}

\author{P.~Makk, Z.~Balogh, Sz.~Csonka, A.~Halbritter}
\affiliation{Department of Physics, Budapest University of
Technology and Economics and \\
Condensed Matter Research Group of the Hungarian Academy of Sciences, 1111 Budapest, Budafoki ut 8., Hungary}

\date{\today}

\begin{abstract}

The interaction of carbon monoxide molecules with atomic-scale platinum nanojunctions is investigated by low temperature mechanically controllable break junction experiments. Combining plateaus' length analysis, two dimensional conductance-displacement histograms and conditional correlation analysis a comprehensive microscopic picture is proposed about the formation and evolution of Pt-CO-Pt single-molecule configurations. Our analysis implies that before pure Pt monoatomic chains would be formed a CO molecule infiltrates the junction, first in a configuration being perpendicular to the contact axis. This molecular junction is strong enough to pull a monoatomic platinum chain with the molecule being incorporated in the chain. Along the chain formation the molecule can either stay in the perpendicular configuration, or rotate to a parallel configuration. The evolution of the single-molecule configurations along the junction displacement shows quantitative agreement with theoretical predictions, justifying the interpretation in terms of perpendicular and parallel molecular alignment. Our analysis demonstrates that the combination of two dimensional conductance-displacement histograms with conditional correlation analysis is a useful tool to separately analyze fundamentally different types of junction trajectories in single molecule break junction experiments.

\end{abstract}

\maketitle

\section*{Introduction}

It is a great challenge of nanoscience to explore electron transport at the ultimate smallest length-scale, where the
current flows through single atoms or single molecules.\cite{agrait03,molecularelectronics} At atomic dimensions the fabrication of nanostructures strongly relies on the self-organizing properties of matter. It was a striking discovery that nature can even form one dimensional metallic wires with single-atom cross section along the rupture of Au, Pt and Ir nanojunctions.\cite{yanson98,ohnishi98,smit01} Such reduced dimensional structures exhibit an enhanced ability for chemical interactions, giving rise to the formation of rich architectures, like molecule-decorated atomic chains.\cite{PhysRevB.73.075405,PhysRevLett.96.026806,PhysRevLett.98.146802,doi:10.1021/jz100084a,doi:10.1021/nn200759s} A single-molecule junction within an atomic chain is
an ideal test system to explore the future possibility of molecular electronics with atomic-scale interconnection.

Break junction techniques\cite{agrait03} are among the most widely used methods to create atomic-scale and single molecule structures along the controlled mechanical elongation of a metallic wire in molecular environment.\cite{doi:10.1021/ja0762386,doi:10.1021/nl052373+,reed97,smit02,xu03,NanoLett_6_2238_2006, doi:10.1021/ja107340t,doi:10.1021/jp204005v, PhysRevB.73.075405,PhysRevLett.96.026806,PhysRevLett.98.146802,doi:10.1021/jz100084a,doi:10.1021/nn200759s} The atomic resolution imaging of such structures is highly demanding,\cite{kondo00,ohnishi98,doi:10.1021/nl049725h,rodrigues00} thus the microscopic behavior is usually traced back from indirect conductance measurements and their comparison with ab initio simulations.\cite{PhysRevLett.88.256803, PhysRevLett.96.046803, Mischenko_Nanolett_10_156_2010, doi:10.1021/ja103327f,indium} In the most common experiment the evolution of the conductance is monitored during the repeated opening and closing of the nanojunction, and the captured conductance vs.\ electrode-separation traces are analyzed by conductance histograms. Peaks in the histogram reflect the conductance of stable, frequently occurring junction configurations. It is, however, obvious that conductance histograms supply a very limited information about the system under study. To gain more information one can explore properties beyond the linear conductance. The measurement of shot noise\cite{PhysRevLett.82.1526,doi:10.1021/nl060116e,doi:10.1021/nl904052r}, conductance fluctuations,\cite{ludoph99,PhysRevB.69.121411} subgap structures in superconducting contacts\cite{scheer98,makk08} or thermopower\cite{ludoph99a, PhysRevB.70.195107} supply useful information about the elastic transmission properties of the junction, whereas point contact spectroscopy\cite{smit02,PhysRevLett.100.196804} or inelastic electron tunneling spectroscopy\cite{PhysRevB.81.075405,doi:10.1021/nn200759s} is widely used to explore inelastic degrees of freedom, like the excitation of single-molecule vibration modes. A further possibility to access information beyond conductance histograms is the advanced statistical analysis of conductance versus electrode separation traces.\cite{Martin_JACS_130_13198_2008,Quek2009, PhysRevLett.105.266805,doi:10.1021/nn300440f} This approach is especially popular in room temperature single-molecule conductance measurements, where the above advanced measurement techniques are usually not applicable due to the reduced energy resolution and stability. In this paper we demonstrate a novel combination of recently introduced statistical analysis techniques to understand a simple test system, the formation and evolution of carbon monoxide single-molecule junctions along the rupture of Pt nanowires and atomic chains.

Recent Pt-CO-Pt conductance histogram measurements\cite{PhysRevB.80.085427, 0957-4484-18-3-035205} demonstrate the formation of two distinct, well separable molecular binding configurations, however the microscopic nature of these configurations was not yet clarified. Here, we demonstrate a detailed statistical analysis of our experimental Pt-CO-Pt break junction data (see experimental methods for details) based on length analysis of the conductance traces with plateaus' length histograms\cite{yanson98} and two dimensional conductance electrode separation histograms.\cite{Martin_JACS_130_13198_2008,Quek2009} To separate several fundamentally different junction evolution trajectories we combine the length analysis with a conditional correlation analysis of the traces, a method recently introduced by the authors.\cite{PhysRevLett.105.266805,doi:10.1021/nn300440f} This combined analysis allows us to resolve fine microscopic details of atomic-scale junction formation and evolution which are hidden in conductance histograms, and obscured in mere length analysis methods. Our results imply that \emph{before} pure Pt monoatomic chains would be formed a CO molecule infiltrates the junction, first in a configuration being perpendicular to the contact axis. This molecular junction is strong enough to pull a Pt atomic chain with the molecule being incorporated in the chain. Along the chain formation the molecule can either stay in the perpendicular configuration, or rotate to a parallel configuration.
The proposed evolution of the single-molecule configurations along the junction displacement is in quantitative agreement with theoretical simulations.\cite{PhysRevB.73.125424} Our analysis also demonstrates that conditional correlation analysis may serve a more complete understanding for any single molecule system, where multiple fundamentally different junction configurations are mixed in the conductance data.

\section*{Results and Discussion}

\subsection*{Conductance histograms}

\begin{figure}[!htb]
\begin{center}
\includegraphics[width=\columnwidth]{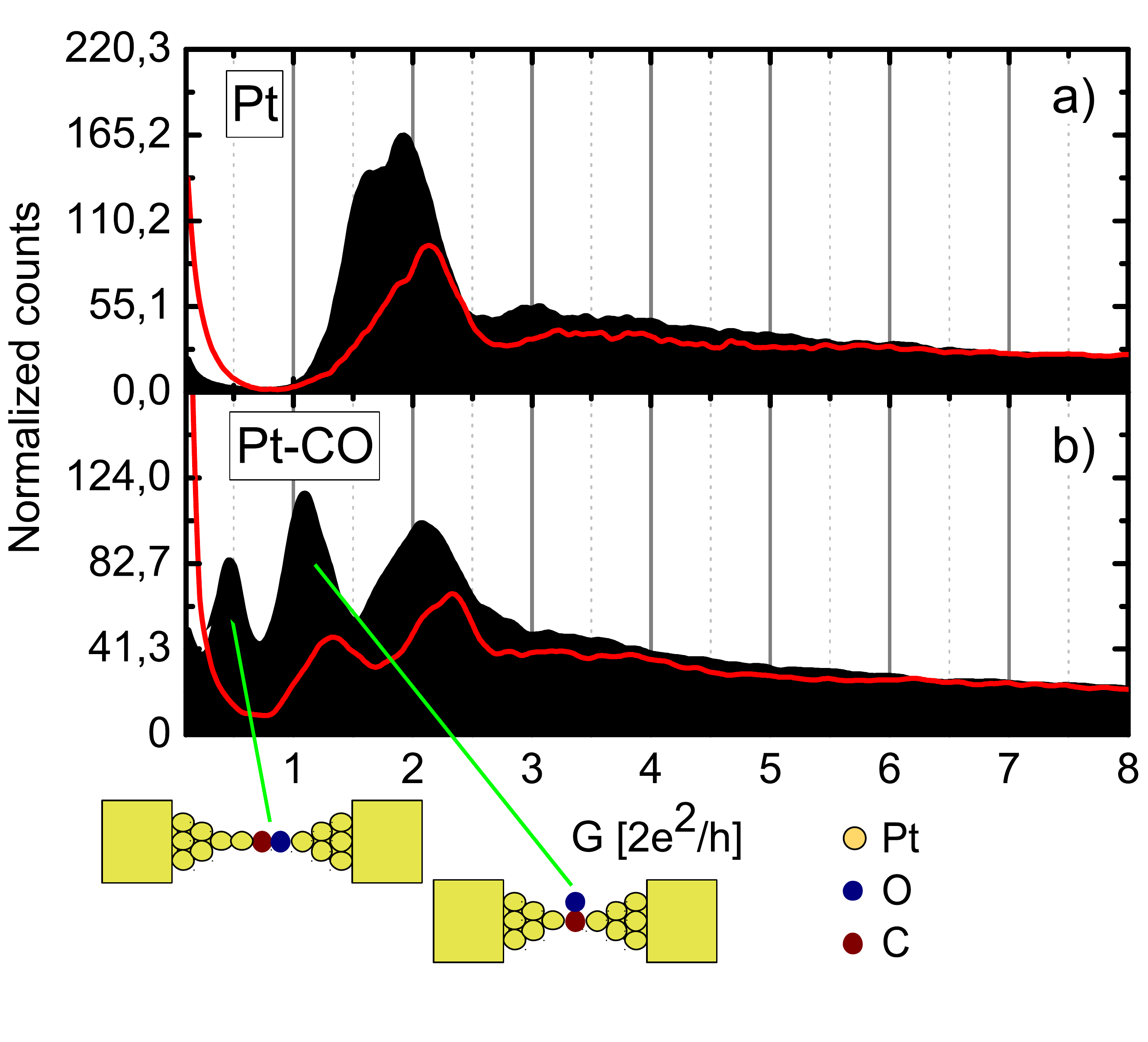}
\end{center}
\caption{\emph{a: Conductance histogram of pure Pt nanocontacts constructed from pulling (shaded black) and pushing traces (red). b: Pulling (shaded black) and pushing (red) conductance histogram of Pt in CO environment. The cartoons show the configurations identified with the histogram peaks.}} \label{tiszta_vs_molekula.eps}
\end{figure}

We start our analysis with the demonstration of conductance histograms, reproducing the results of Tal et al.\cite{PhysRevB.80.085427, 0957-4484-18-3-035205} Fig.~\ref{tiszta_vs_molekula.eps}b shows the typical Pt-CO conductance histogram both for the traces that were recorded along the opening of the junction (\emph{pulling histogram}, see shaded black graph) and along the closing of the junction (\emph{pushing histogram}, see red line). As a reference, Fig.~\ref{tiszta_vs_molekula.eps}a demonstrates the pulling and pushing histogram for pure Pt junctions without CO molecules.

The pure Pt pulling histogram shows a peak at $\approx 1.9$~G$_0$ with a small shoulder at $\approx 1.6$~G$_0$, which is typical for low temperature measurements.\cite{agrait03, PhysRevB.80.085427, 0957-4484-18-3-035205, nielsen03} As CO is dosed to the junction two new peaks appear at $0.5$ and $1.1\,$G$_0$ and the single-atom peak of pure Pt shifts to $2.1$~G$_0$ conductance.\cite{PhysRevB.80.085427, 0957-4484-18-3-035205} In pure Pt the first peak in the pushing histogram is significantly smaller and has higher conductance than the one in the pulling histogram. For molecular junctions the peak at $0.5$~G$_0$ completely disappears in the pushing histogram, and the other two peaks become smaller and slightly shift towards higher conductance compared to the pulling histogram.

To identify the molecular geometries the ab initio calculations of Strange et al.\ are used.\cite{PhysRevB.73.125424}  The authors have calculated the conductance of a Pt-CO-Pt junction, where a CO molecule has been inserted to a dimer Pt junction with the carbon atom binding to the electrodes, and the oxygen atom hanging out perpendicular to the contact axis (see right cartoon in Fig.~\ref{tiszta_vs_molekula.eps}). As the electrodes are pulled apart the molecule rotates to a parallel configuration (see left cartoon in Fig.~\ref{tiszta_vs_molekula.eps}), where both the C and O atoms lie in the contact axis. As a reference a pure Pt dimer configuration and a pure Pt chain with 3 atoms was also investigated. For the perpendicular and parallel molecular configurations the calculations have revealed conductances of $1.5\,$G$_0$ and $0.5\,$G$_0$, respectively.
This suggests, that the lower peak in the conductance histogram of Fig.~\ref{tiszta_vs_molekula.eps}b corresponds to a parallel configuration of the CO molecule. The calculated $1.5\,$G$_0$ conductance of the perpendicular configuration deviates from the position of the the second histogram peak ($1.1\,$G$_0$). Still, we use the working hypothesis that the $1.1\,$G$_0$ peak reflects a perpendicular molecular configuration, which is later supported by experimental observations. The histogram peak at $2.1\,$G$_0$ is attributed to pure Pt junctions.

In Figure~\ref{tiszta_vs_molekula.eps}a a significant difference is observed between the pushing and pulling histogram of pure Pt, which is attributed to the formation of monoatomic chains.\cite{nielsen03,smit01} Along the disconnection of the junction atomic chains are pulled, which are reflected by a long plateau introducing a large weight to the first peak of the pulling histogram. After the complete rupture the chain atoms fall back to the electrodes, and along the closing of the junction only a much shorter single-atom plateau is observed.
Furthermore, along the chain formation the conductance slightly decreases compared to the single-atom conductance,\cite{nielsen03} whereas along the closing of the junction only the single atom conductance is observed. All these explain the higher position and the smaller weight of the peak in the pushing histogram.
The clear difference between the pulling and pushing Pt-CO histogram may also be associated with a chain formation process. Next this possibility is analyzed.

\subsection*{Plateaus' length analysis}

\begin{figure}[!htb]
\begin{center}
\includegraphics[width=\columnwidth]{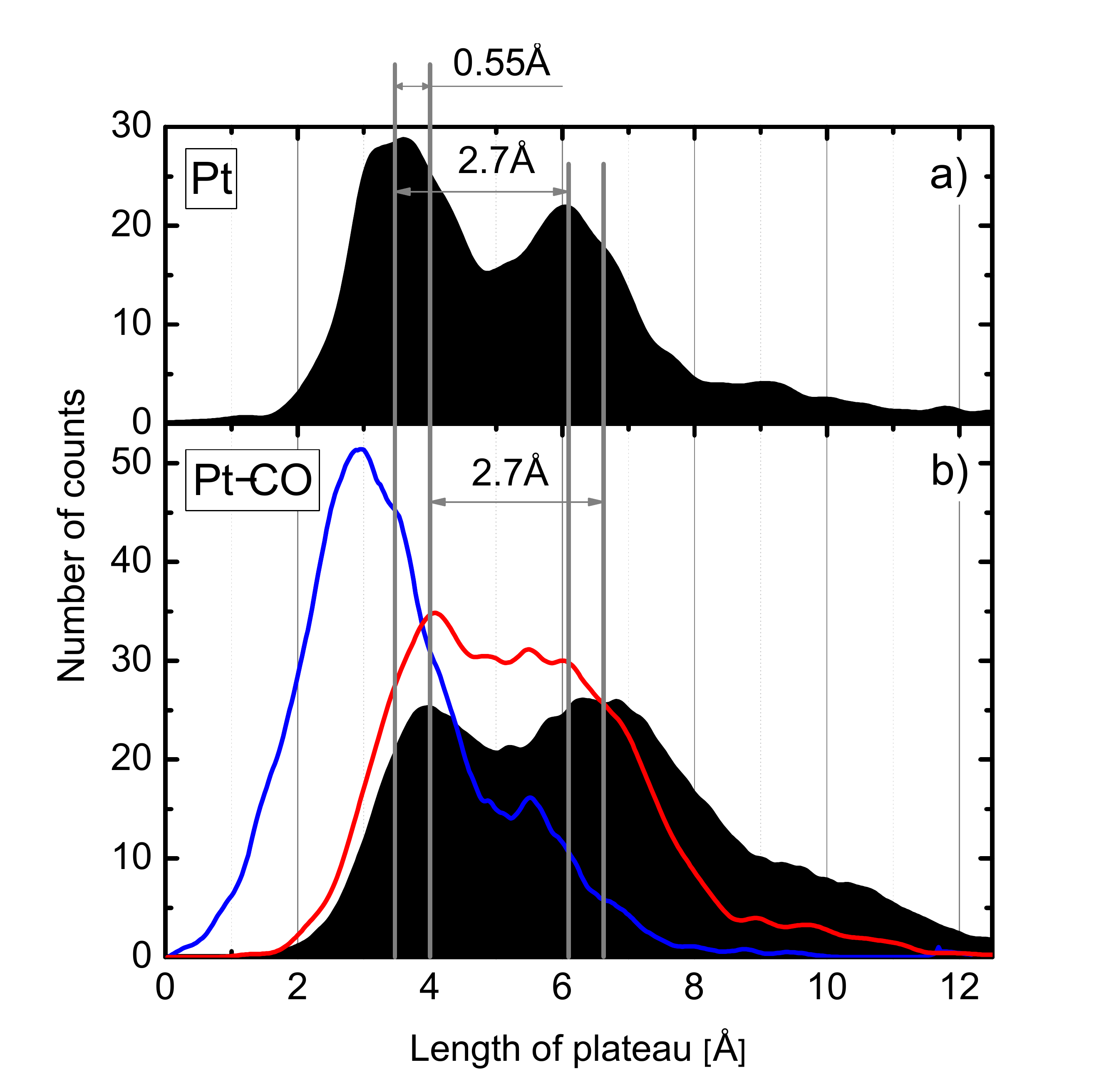}
\end{center}
\caption{\emph{(a): PLH for pure Pt nanojunctions constructed for the $0.2-2.75\,$G$_0$ conductance interval. The distance of the PLH peaks has been set to $2.7$\AA . (b): PLHs for Pt-CO junctions, using different conductance intervals to measure the plateaus' length (see text).}} \label{pt_vs_CO_platoh.eps}
\end{figure}

Figure~\ref{pt_vs_CO_platoh.eps}a shows the plateaus' length histogram\cite{yanson98} (PLH) constructed for pure platinum junctions. This histogram shows how the length of the final conductance plateau is distributed among a large ensemble of traces.
On a particular conductance trace the length of the last plateau is measured as the electrode displacement between the first and last datapoint within the conductance interval of the histogram peak, $0.2-2.75\,$G$_0$.
The observation of equidistant peaks in the PLH has provided a strong evidence for the formation of monoatomic chains:\cite{yanson98} the peaks of the PLH correspond to the rupture of chains with different number of atoms, whereas the separation of neighbor peaks tells the interatomic distance in the chain. Our results agree with previous observations, the chain formation process is clearly resolved by the PLH.
In the following discussions the separation of the PLH peaks is identified with the interatomic Pt-Pt distance of $2.7\,$\AA~ obtained in the calculations of Strange et al.\cite{PhysRevB.73.125424} This choice for the calibration of the electrode separation will enable a direct comparison between measurement and simulation. As the pure Pt and Pt-CO histograms in Figure~\ref{tiszta_vs_molekula.eps} were measured on the same sample, this calibration also applies for the Pt-CO contacts.

We have also constructed the PLH for the molecular measurements. For pure Pt junctions the conductance region, in which the plateaus' length is measured is obviously given by the position of the single peak in the conductance histogram. For molecular junctions, however, several choices are possible. Figure~\ref{pt_vs_CO_platoh.eps}b shows three PLHs created for the molecular junctions. The blue curve shows the PLH constructed for the conductance region of the third histogram peak ($1.4-2.75$~G$_0$). For this conductance interval no sign of chain formation is observed.
If the plateaus' length is measured in the joint interval of the second and third histogram peak ($0.7-2.75$~G$_0$), the PLH already gets wider, but no clear peaks are resolved. However, if the the conductance interval is extended to  all the three conductance histogram peaks ($0.2-2.75$~G$_0$) a clear peak structure is resolved in the PLH indicating a chain formation process.

As the length of the last plateau is measured from the same reference conductance value of $2.75\,$G$_0$ both for pure and molecular junctions, it is possible to compare the pure and molecular PLHs. This comparison shows two apparent features: (i) the separation of the peak in the PLH is similar for pure and molecular junctions, which indicates that the PLH of the molecular junction is also related to the formation of Pt atomic chains with a characteristic interatomic distance of ~$2.7\,$\AA. (ii) For molecular junctions the entire PLH is shifted towards larger displacements with on offset of $0.55\,$\AA~ compared to pure junctions. This offset is naturally interpreted by the infiltration of the CO molecule to the junction \emph{before} the formation of Pt atomic chain, which is further justified by forthcoming analysis.

\subsection*{Two dimensional conductance-displacement histograms}

To give a more complete view of the chain formation process, two dimensional conductance-displacement histograms\cite{Martin_JACS_130_13198_2008,Quek2009,doi:10.1021/nn300440f} (2DCDH) have been constructed as follows.
Each conductance trace represents conductance as a function of electrode displacement, $G(x)$.  To compare different conductance traces, each trace is offset along the distance scale, $x$ such that at the new distance origin all the traces take the same reference conductance value, $G_{\mathrm{ref}}$. Afterwards a two dimensional histogram is constructed from all the data points of the offset traces as a function of both the conductance and the distance, signing the counts of the 2D histogram by color scale. Such 2D conductance-displacement histograms are demonstrated in Fig.~\ref{pt_vs_CO_platoh.eps}a,d both for pure and CO affected Pt junctions using the same data as in Fig.\ref{tiszta_vs_molekula.eps}. As a reference conductance $G_\mathrm{ref} = 2.75\,$G$_0$ was chosen, which coincides with the upper limit of the conductance intervals for which the PLHs were constructed (shaded black graphs in Fig.\ref{pt_vs_CO_platoh.eps}).

\begin{figure}[!htb]
\begin{center}
\includegraphics[width=\columnwidth]{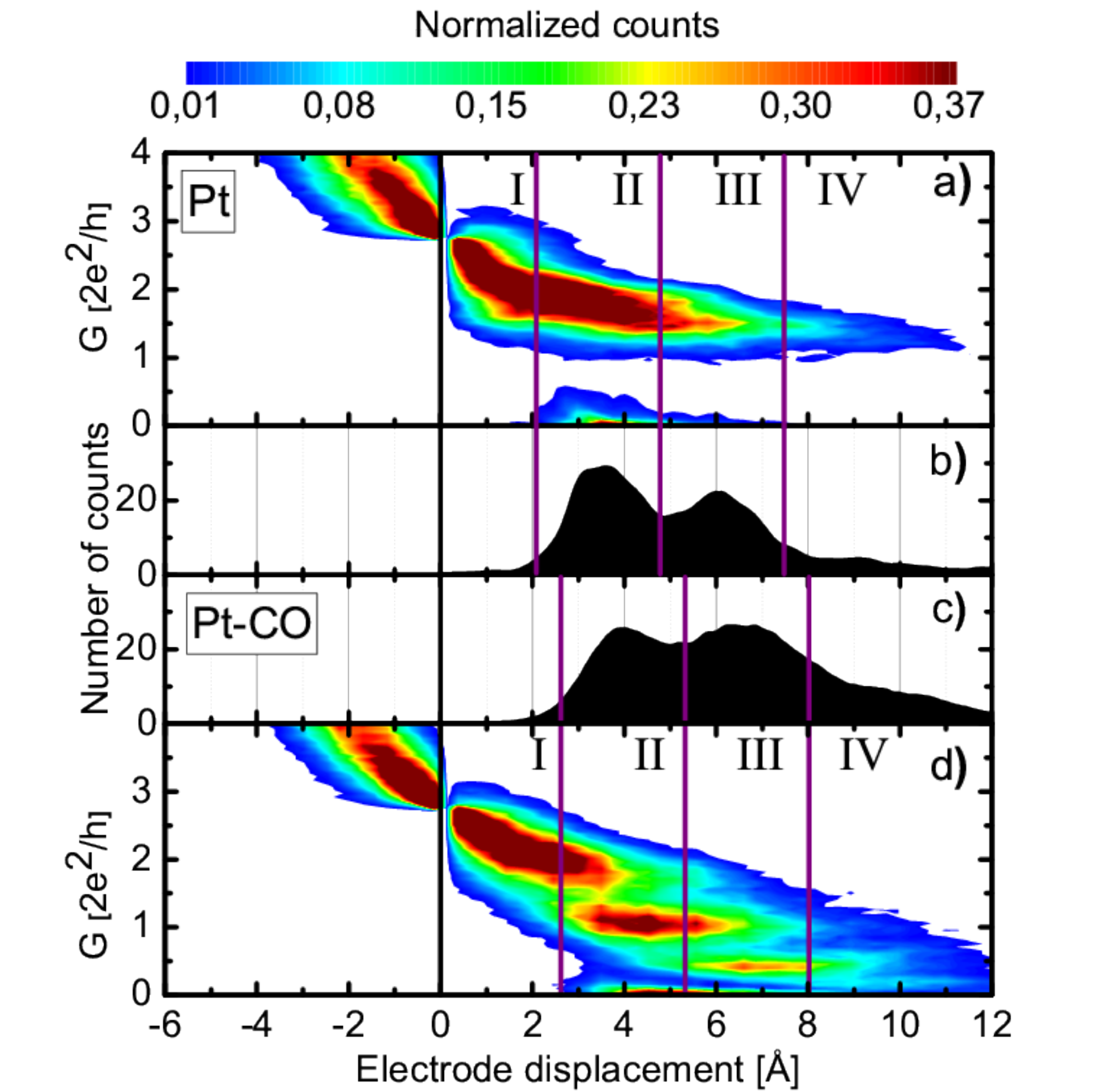}
\end{center}
\caption{\emph{(a,b): 2DCDH for pure Pt nanojunctions and the corresponding PLH, respectively. (c,d): PLH and 2DCDH  for Pt-CO junctions, respectively. The colorscale at the top represents the counts of both 2DCDHs normalized to the number of included conductance traces.}} \label{pt_vs_CO_2Dtrace.eps}
\end{figure}

Figure~\ref{pt_vs_CO_2Dtrace.eps}a shows the 2DCDH for pure platinum junctions. The 2D histogram has been divided into four displacement regions (I-IV) separated by purple lines in the Figure. The middle line is fixed at the minimum between the two peaks of the corresponding PLH (Fig.~\ref{pt_vs_CO_2Dtrace.eps}b), which separates the regions of monoatomic contacts (marked as I-II) from  chain forming regions (marked as III-IV). The two additional lines have been drawn such, that the width of regions II and III has been set to the platinum interatomic distance, $2.7\,$\AA. With
this choice the four regions are interpreted as follows. Region I corresponds to displacements where monoatomic contacts are already formed, but the rupture of the contact is unlikely (as reflected by the low counts of the PLH). Region II corresponds to the displacement interval where the monoatomic contacts usually break. Region III corresponds to the interval where already one atom is pulled to the atomic chain. Finally, region IV marks the interval, where already more than one atoms are pulled to the atomic chain.

The 2DCDH of pure Pt in Fig.~\ref{pt_vs_CO_2Dtrace.eps}a shows a single pronounced plateau in the conductance range of the histogram peak which also extends to regions III-IV as a result of atomic chain formation. It is also seen, that the conductance of this configuration decreases along the chain formation process.

The 2DCDH for the molecular measurements in Fig.~\ref{pt_vs_CO_2Dtrace.eps}d clearly shows the appearance of the two new molecular configurations at $1.1$ and $0.5\,$G$_0$ conductance. Here, the traces have also been separated into four regions (I-IV), but now the PLH in panel (c) corresponding to the molecular measurements has defined the regions (see purple lines). The width of region II and III has also been set to $2.7\,$\AA, and the PLH in panel (c) is the same as the one in Figure~\ref{pt_vs_CO_platoh.eps}b calculated for the $0.2-2.75\,$G$_0$ conductance region.

The first region (I) of the 2D histogram -- where the contact is not likely to break -- is clearly dominated by single-atom configurations with $\sim 1.4-2.75\,$G$_0$ conductance. Region II is dominated by the $1.1\,$G$_0$ configuration, but the $0.5\,$G$_0$ configuration does not yet appear in this region. The regions where chains are formed (III and IV) are mainly composed of configurations with either $1.1$ or $0.5\,$G$_0$ conductance.

At this point we find it useful to emphasize the fundamental difference between PLHs and 2DCDHs. Whereas the horizontal scale of the PLH corresponds to the distance at which the junction finally \emph{breaks}, the 2DCDH contains all the datapoints before the final rupture of the junction. As an example region II in the 2DCDH does not only represent the traces breaking in this displacement interval, but also the longer traces, broken in regions III-IV, give contribution to region II. This means that the 2DCDH mixes the information about fundamentally different types of traces. To separate this mixed information we perform \emph{a conditional analysis}, i.e. we construct 2DCDHs, conductance histograms and PLHs for selected subsets of the traces that fulfill a particular condition. First we separately analyze the 2DCDHs for those traces that break in regions II, III or IV, respectively. As a next step we separate the traces according to the nature of the final configuration before rupture.

\subsection*{Conditional analysis according to the breaking length}

\begin{figure}[!htb]
\begin{center}
\includegraphics[width=\columnwidth]{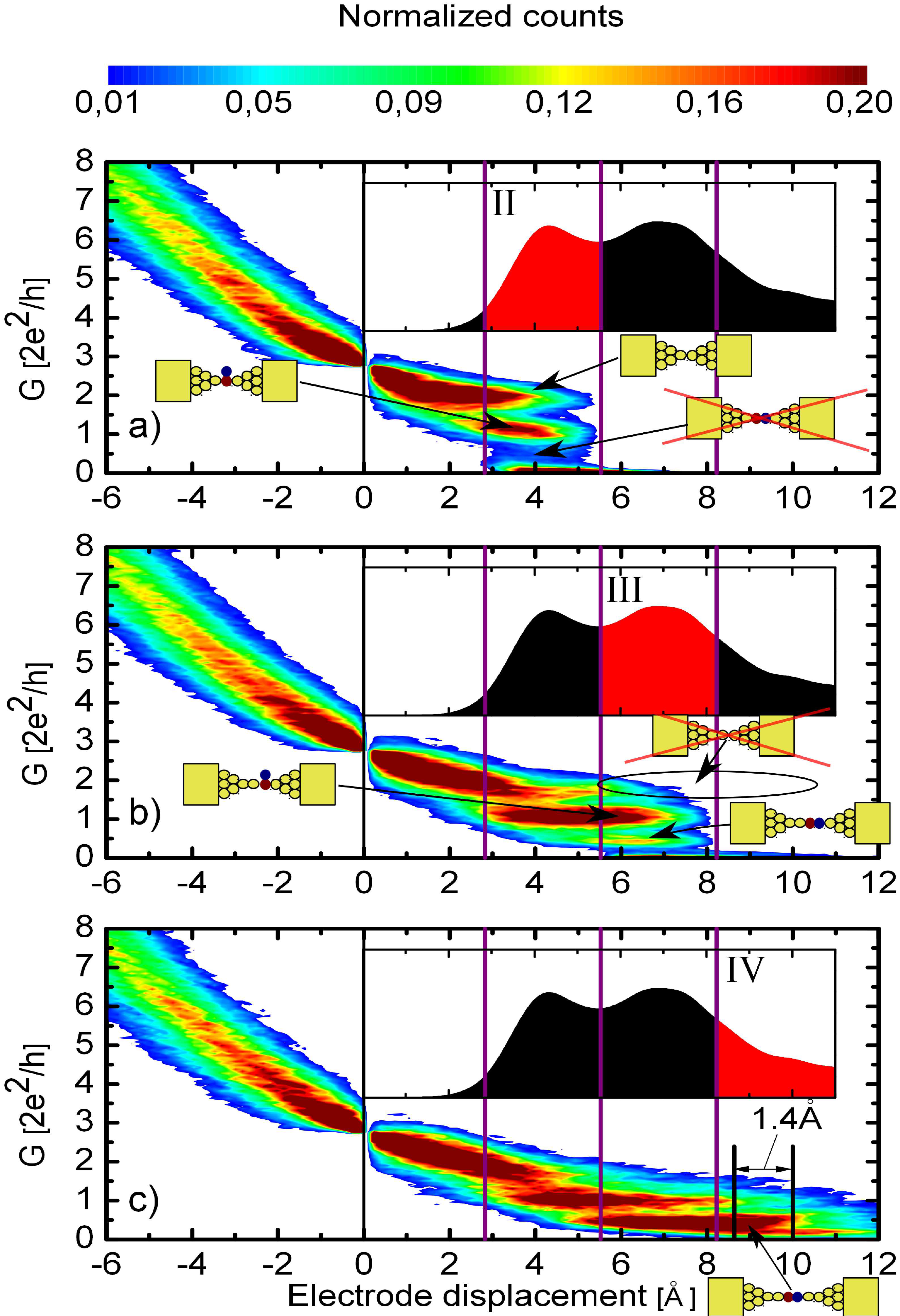}
\end{center}
\caption{\emph{2DCDHs for those selected traces that break in region II (a), region III (b). or in region IV (c) of the PLH. As a reference the PLH for the whole Pt-CO dataset is given in the insets using the same electrode displacement scale as that of the 2DCDHs. The red regions of the PLHs mark the displacement regions for which the selection is performed. The cartoons are illustrations for the possible junction geometries.}} \label{2D_int1_wide.eps}
\end{figure}

In Figure~\ref{2D_int1_wide.eps}a we demonstrate the 2DCDH for those selected \emph{traces, which break without chain formation}, i.e.\ which correspond to region II of the PLH. In comparison to Fig.~\ref{pt_vs_CO_2Dtrace.eps} it can be noticed that the $0.5\,$G$_0$ configuration is completely suppressed in the 2DCDH, and the weight of the $1.1\,$G$_0$ configuration is also decreased. This means that the parallel configuration of the molecule only appears if chains are pulled.

As a next step we analyze traces that break after one atom has been pulled to the chain, i.e. which correspond to region III of the PLH. Fig.~\ref{2D_int1_wide.eps}b demonstrates 2DCDH for these selected traces. Here the weight of the $2.1\,$G$_0$ configuration is negligible in region III of the 2DCDH demonstrating that the formation of pure Pt atomic chains is not likely. The strong weight of the $1.1\,$G$_0$ configuration spans both region II and III. This shows that the perpendicular molecular configurations binds to the junction before a chain would be formed, and this single-molecule junction is strong enough to pull a Pt atomic chain out of the electrodes. The $0.5\,$G$_0$ configuration only appears in region III, i.e. the parallel configuration is only formed during the chain formation process and not before.

Finally, we investigate those traces that break after pulling more than one atom to the chain. For these traces
the $1.1\,$G$_0$ configuration extends to region IV, showing that the perpendicular configuration can pull chains with more than one atoms. The $0.5\,$G$_0$ configurations shows an extended plateau in regions III and IV spanning a displacement interval of $\approx 5\,$\AA. This indicates that the the rotation of the molecule to the parallel configuration can occur well before the rupture of the chain, i.e. the parallel configuration also binds strongly enough to the electrodes to pull Pt atomic chains.\cite{note}

\subsection*{Conditional analysis according to the nature of the final configuration}

So far the traces were classified based on the breaking distance. It is also useful to sort the traces according to the nature of the final configuration before rupture. A trace is considered to break from one of the three contact configurations ($0.5$, $1.1$ or $2.1\,$G$_0$) if $75\,\% $ of the final $0.5\,$\AA~ displacement before rupture is within the conductance region of the corresponding histogram peak. With this selection criterium the traces can be sorted to distinct sets according to the final configuration, such that $91\%$ of all traces fulfill the above criteria for one of the three conductance intervals. For the the remaining $9\%$ of the traces the final configuration is not clearly recognizable due to the scattering of the conductance data within multiple peak regions, so these traces are rejected in the following analysis. This final configuration analysis reveals quantitative information about the rate of various, fundamentally different types of ruptures.

It is found that in $74\,\% $ of all traces a CO molecule binds to the junction. From these $44\,\% $ of all traces break from the perpendicular configuration, and the remaining $30\,\% $ break from the parallel configuration (see Table~\ref{table_corr}). The conditional conductance histogram (Fig.~\ref{fig_supp.eps}a) constructed from the former set of traces does not show any peak at $0.5\,$G$_0$, i.e. if the junction breaks from a perpendicular configuration then a parallel configuration does not appear before. The conditional conductance histogram for junctions breaking from the parallel configuration exhibits a clear peak at $1.1\,$G$_0$ with the same height as the total histogram. This shows that the parallel configuration is preceded by a perpendicular configuration.

\begin{table}[!htb]
\begin{center}
\begin{tabular}{|c|| c | c | c| c| c| }
\hline
  & $\mathbf{Region\,I}$ & $\mathbf{Region\,II}$ & $\mathbf{Region\,III}$
& $\mathbf{Region\,IV}$& \textbf{Sum}\\ \hline \hline
{\bf $\mathbf{2.1\,}$G$\mathbf{_0}$}  & $0\%$ & $9\%$ & $8\%$ & $0\%$ & $17\%$\\ \hline
{\bf $\mathbf{1.1\,}$G$\mathbf{_0}$}  & $0\%$ & $12\%$ & $23\%$ & $9\%$ & $44\%$\\ \hline
{\bf $\mathbf{0.5\,}$G$\mathbf{_0}$}   & $0\%$ & $4\%$ & $9\%$ & $17\%$ & $30\%$\\ \hline
\textbf{Sum}   & $0\%$ & $25\%$ & $40\%$ & $26\%$ & 91\%\\ \hline
\end{tabular}
\caption{\emph{The number of the conductance traces breaking from a certain final configuration (rows) within a certain displacement interval (columns) relative to the total number of traces. The final row (column) gives the sum of the preceding rows (columns). All the numbers are rounded to integer percentage values.}}
\label{table_corr}
\end{center}
\end{table}

Beside the nature of the final configuration the traces can be further classified according to the breaking distance (I-IV regions) as demonstrated by Table~\ref{table_corr}. It is found that in $58\,\% $ of the traces a Pt chain is pulled by the CO molecule ($1.1$ or $0.5\,$G$_0$ final configuration and region III or IV), from these in $26\,\% $ of all traces the molecule finally rotates to a parallel configuration, and in the remaining $32\,\% $ of all traces the junction breaks from the perpendicular molecular configuration. It is also clear that for long chains (region IV) the formation of a parallel configuration is more likely than the rupture from the perpendicular configuration, whereas for short chains (region III) the opposite is observed. As a comparison, the rate of pure Pt chains is only $8\,\% $ ($2.1\,$G$_0$ final configuration and region III or IV).

The classification of final configurations is also useful to perform a more detailed plateaus' length analysis, i.e.\ to plot the PLHs separately for the traces breaking from a certain final configuration. The orange and blue PLHs in Fig.~\ref{fig_supp.eps}b are plotted for the traces that break from the $1.1\,$G$_0$ or the $0.5\,$G$_0$ configuration, respectively. As reference the shaded grey graph represents the PLH for the whole Pt-CO dataset, whereas the green line shows the PLH for pure Pt junctions (the same as the shaded black PLHs in Fig.~\ref{pt_vs_CO_platoh.eps}). This analysis shows two important features:

\begin{figure}[!htb]
\begin{center}
\includegraphics[width=\columnwidth]{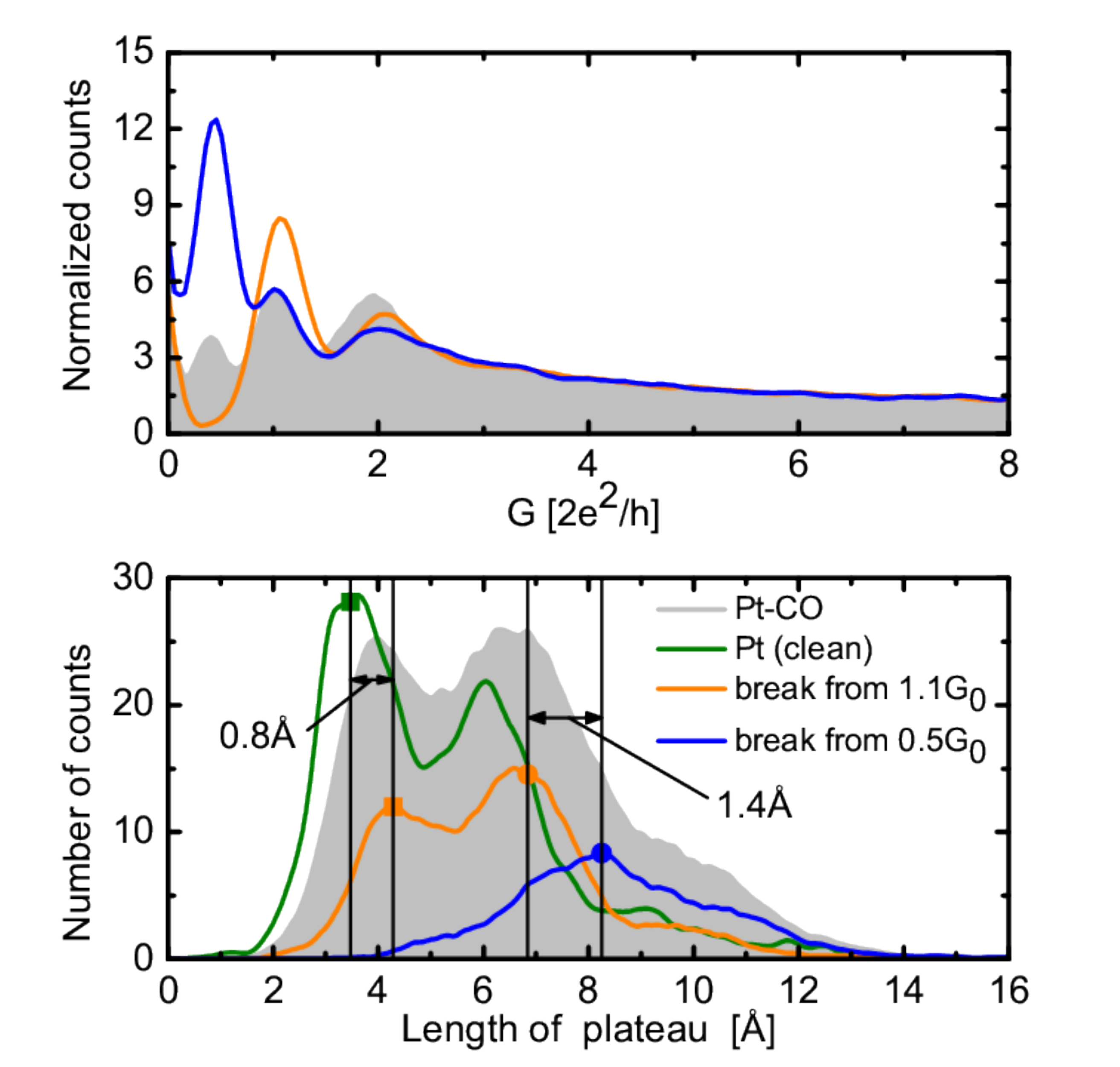}
\end{center}
\caption{\emph{(a): Conditional conductance histograms, i.e. conductance histograms for those selected traces that break from the $1.1\,$G$_0$ configuration (orange line) or the $0.5\,$G$_0$ configuration (blue line). As a reference the shaded grey graph shows the conductance histogram for the whole dataset (same as the shaded black graph in Fig.~\ref{tiszta_vs_molekula.eps}a). All histograms are normalized to the number of included traces. (b): Conditional PLHs for those traces that break from the $1.1\,$G$_0$ configuration (orange line) or the $0.5\,$G$_0$ configuration (blue line). As reference the shaded grey graph and the green line show the PLH for the whole Pt-CO dataset and the PLH for for pure Pt junctions, respectively.}} \label{fig_supp.eps}
\end{figure}

(i): The peak of the PLH for traces breaking from the parallel configuration is shifted by $\approx 1.4\,$\AA~ compared to the PLH of the traces breaking from the perpendicular configuration. This agrees with the calculated electrode displacement of $\approx 1.1\,$\AA~\cite{PhysRevB.73.125424} (practically the C-O length) which is required to rotate the perpendicular molecule to a parallel configuration. (Note, that similar information is also revealed by the 2DCDH in Fig. \ref{2D_int1_wide.eps}c, where the $0.5\,$G$_0$ configuration extends to $\approx1.4\,$\AA~ longer displacement than the $1.1\,$G$_0$ configuration)

(ii): The offset of the orange PLH peaks (traces breaking from the $1.1\,$G$_0$ configuration) with respect to the pure Pt PLH peaks (green line) is $\approx 0.8\,$\AA~. This displacement is in perfect agreement with the calculated elongation of $0.75\,$\AA~,\cite{PhysRevB.73.125424} which is required for the incorporation of a perpendicular CO molecule in the Pt atomic contact. It is worth emphasizing, that the PLH peaks of the total Pt-CO dataset show smaller offset with respect to the pure Pt PLH peaks ($\approx 0.55\,$\AA~, see Fig.~\ref{pt_vs_CO_platoh.eps}), which is attributed to the fact that about $17\,\% $ of the traces in the total Pt-CO dataset do not show any molecular configurations, thus this PLH mixes the features of pure and molecular junctions.

These two observations provide strong quantitative arguments that the description of the two molecular configurations by a perpendicular and a parallel arrangement of the CO molecule is indeed reasonable. The deviation between the measured and calculated conductance of the perpendicular configuration ($1.1$ and $1.5\,$G$_0$) may arise from the idealized nature\cite{indium} of the simulations in Ref.~\onlinecite{PhysRevB.73.125424}. Either the number of nearest neighbor atoms in the electrodes or the constraint of placing the Pt dimer atoms at the contact axis may deviate from the typical experimental situation. The long chain-like parallel configuration is naturally less sensitive to the fine details of contact geometry, and thus the calculated conductance is more accurate.

\section*{Conclusions}

In conclusions we propose a consistent microscopic picture about of the formation and evolution of CO single-molecule junctions contacted by Pt electrodes based on a comprehensive conditional analysis of the conductance traces: In the majority of the cases the CO molecule is incorporated into the Pt atomic junction, first in a perpendicular configuration, which may later rotate to a parallel configuration. This perpendicular molecular configuration is established before the formation of Pt chains, i.e. the formation of a pure Pt atomic chain is not likely. Our results demonstrate that both the perpendicular and the parallel configuration is strong enough to pull Pt atomic chains. The molecule does not always rotate to a parallel final configuration, but the probability of this process increases with the chain length. The analysis of the data reveals two characteristic displacements: (i) the displacements at which a perpendicular CO molecule can be incorporated in the Pt atomic junction; (ii) the displacement required to rotate the perpendicular molecular configuration to a parallel one. These measured displacement values agree with the results of computer simulations, which supports our interpretation in terms of perpendicular and parallel configurations.
Our results also demonstrate that the combination of \emph{conditional correlation analysis}\cite{doi:10.1021/nn300440f} with two dimensional conductance-displacement and plateaus' length histograms provides a powerful method to separately resolve divers junction evolution trajectories in single-molecule break junction data.

\section*{Experimental methods}

All experiments were performed by conventional mechanically controllable break junction technique\cite{agrait03} at $T=4.2\,$K temperature. The sample was prepared from a $100\,\mu$m diameter high purity Pt wire, which was glued by two drops of Stycast epoxy to a phosphor-bronze bending beam. Between the fixing points a notch was cut in the wire with a razor blade reducing the diameter to $\approx 10\,\mu$m. The sample was broken in situ in the cryogenic vacuum with a mechanical actuator. Afterwards a piezo actuator was used for the fine tuning of the electrode separation. Conductance histograms were measured by repeatedly opening and closing the junction with a symmetric triangle signal on the piezo (10Hz). The conductance was measured in a voltage biased setup with $V=100\,$mV bias. The current through the junction was amplified with a FEMTO DLPCA-200 current amplifier, and the amplified signal was acquired by a National Instruments data acquisition card (100kS/s sampling). The conductance data were were separated to pulling and pushing traces, from which the pulling and pushing histograms were constructed. All the measurements were based on the analysis of $>5000$ conductance traces.

The CO molecules were dosed to the junction through stainless steel tube going from a room temperature vacuum flange to the vicinity of the cryogenic temperature sample. This tube was heated above the boiling point of CO by a $0.5\,$mm diameter \emph{thermocoax} wire which was rolled helically inside the stainless steel tube. To prevent unwanted contamination the exit of the stainless steel tube was isolated from the sample by an electromagnetically actuated shutter, which was only opened when the CO molecules were dosed. To eliminate the presence of hydrogen molecules -- which have finite vapor pressure at $4.2\,$K -- a zeolithe sorption pump was placed to the sample holder head, close to the sample. At room temperature the molecules were dosed from a high purity container through a vacuum system evacuated by a turbomolecular pump. Each dosing cycle was preceded by a \emph{fake} dosing cycle imitating all the steps of molecule dosing, but leaving the CO container closed, and checking whether the conductance histogram of pure Pt remains. These imitated cycles are important to make sure that no other molecules than CO may arrive to the junction.

\section*{Acknowledgements}
This work has been supported by the Hungarian research funds OTKA K76010, CNK80991, T\'AMOP - 4.2.2.B-10/1--2010-0009,  EU ERG NanoQuantumDevices 239223. A.H.\ \& S.C\ are grantees of the Bolyai J\'anos Scholarship. The authors acknowledge useful discussions with O.~Tal and J. van Ruitenbeek.
\providecommand*\mcitethebibliography{\thebibliography}
\csname @ifundefined\endcsname{endmcitethebibliography}
  {\let\endmcitethebibliography\endthebibliography}{}


\begin{thebibliography}{0}%
\makeatletter
\providecommand \@ifxundefined [1]{%
 \@ifx{#1\undefined}
}%
\providecommand \@ifnum [1]{%
 \ifnum #1\expandafter \@firstoftwo
 \else \expandafter \@secondoftwo
 \fi
}%
\providecommand \@ifx [1]{%
 \ifx #1\expandafter \@firstoftwo
 \else \expandafter \@secondoftwo
 \fi
}%
\providecommand \natexlab [1]{#1}%
\providecommand \enquote  [1]{``#1''}%
\providecommand \bibnamefont  [1]{#1}%
\providecommand \bibfnamefont [1]{#1}%
\providecommand \citenamefont [1]{#1}%
\providecommand \href@noop [0]{\@secondoftwo}%
\providecommand \href [0]{\begingroup \@sanitize@url \@href}%
\providecommand \@href[1]{\@@startlink{#1}\@@href}%
\providecommand \@@href[1]{\endgroup#1\@@endlink}%
\providecommand \@sanitize@url [0]{\catcode `\\12\catcode `\$12\catcode
  `\&12\catcode `\#12\catcode `\^12\catcode `\_12\catcode `\%12\relax}%
\providecommand \@@startlink[1]{}%
\providecommand \@@endlink[0]{}%
\providecommand \url  [0]{\begingroup\@sanitize@url \@url }%
\providecommand \@url [1]{\endgroup\@href {#1}{\urlprefix }}%
\providecommand \urlprefix  [0]{URL }%
\providecommand \Eprint [0]{\href }%
\providecommand \doibase [0]{http://dx.doi.org/}%
\providecommand \selectlanguage [0]{\@gobble}%
\providecommand \bibinfo  [0]{\@secondoftwo}%
\providecommand \bibfield  [0]{\@secondoftwo}%
\providecommand \translation [1]{[#1]}%
\providecommand \BibitemOpen [0]{}%
\providecommand \bibitemStop [0]{}%
\providecommand \bibitemNoStop [0]{.\EOS\space}%
\providecommand \EOS [0]{\spacefactor3000\relax}%
\providecommand \BibitemShut  [1]{\csname bibitem#1\endcsname}%
\let\auto@bib@innerbib\@empty
\end{thebibliography}%


\begin{mcitethebibliography}{46}
\providecommand*\natexlab[1]{#1}
\providecommand*\mciteSetBstSublistMode[1]{}
\providecommand*\mciteSetBstMaxWidthForm[2]{}
\providecommand*\mciteBstWouldAddEndPuncttrue
  {\def\EndOfBibitem{\unskip.}}
\providecommand*\mciteBstWouldAddEndPunctfalse
  {\let\EndOfBibitem\relax}
\providecommand*\mciteSetBstMidEndSepPunct[3]{}
\providecommand*\mciteSetBstSublistLabelBeginEnd[3]{}
\providecommand*\EndOfBibitem{}
\mciteSetBstSublistMode{f}
\mciteSetBstMaxWidthForm{subitem}{(\alph{mcitesubitemcount})}
\mciteSetBstSublistLabelBeginEnd
  {\mcitemaxwidthsubitemform\space}
  {\relax}
  {\relax}

\bibitem[Agra{\"\i}t et~al.(2003)Agra{\"\i}t, {Levy Yeyati}, and van
  Ruitenbeek]{agrait03}
Agra{\"\i}t,~N.; {Levy Yeyati},~A.; van Ruitenbeek,~J.~M. \emph{Phys. Rep.}
  \textbf{2003}, \emph{377}, 81--279\relax
\mciteBstWouldAddEndPuncttrue
\mciteSetBstMidEndSepPunct{\mcitedefaultmidpunct}
{\mcitedefaultendpunct}{\mcitedefaultseppunct}\relax
\EndOfBibitem
\bibitem[Cuevas and Scheer(2010)Cuevas, and Scheer]{molecularelectronics}
Cuevas,~J.~C.; Scheer,~E. In \emph{Molecular Electronics An introduction to
  Theory and Experiment}; Reed,~M., Ed.; World Scientific: Singapore,
  2010\relax
\mciteBstWouldAddEndPuncttrue
\mciteSetBstMidEndSepPunct{\mcitedefaultmidpunct}
{\mcitedefaultendpunct}{\mcitedefaultseppunct}\relax
\EndOfBibitem
\bibitem[Yanson et~al.(1998)Yanson, {Rubio Bollinger}, van~den Brom,
  Agra{\"{\i}}t, and van Ruitenbeek]{yanson98}
Yanson,~A.~I.; {Rubio Bollinger},~G.; van~den Brom,~H.~E.; Agra{\"{\i}}t,~N.;
  van Ruitenbeek,~J.~M. \emph{Nature} \textbf{1998}, \emph{395}, 783--785\relax
\mciteBstWouldAddEndPuncttrue
\mciteSetBstMidEndSepPunct{\mcitedefaultmidpunct}
{\mcitedefaultendpunct}{\mcitedefaultseppunct}\relax
\EndOfBibitem
\bibitem[Ohnishi et~al.(1998)Ohnishi, Kondo, and Takayanagi]{ohnishi98}
Ohnishi,~H.; Kondo,~Y.; Takayanagi,~K. \emph{Nature} \textbf{1998}, \emph{395},
  780--785\relax
\mciteBstWouldAddEndPuncttrue
\mciteSetBstMidEndSepPunct{\mcitedefaultmidpunct}
{\mcitedefaultendpunct}{\mcitedefaultseppunct}\relax
\EndOfBibitem
\bibitem[Smit et~al.(2001)Smit, Untiedt, Yanson, and van Ruitenbeek]{smit01}
Smit,~R.~H.~M.; Untiedt,~C.; Yanson,~A.~I.; van Ruitenbeek,~J.~M. \emph{Phys.
  Rev. Lett.} \textbf{2001}, \emph{87}, 266102\relax
\mciteBstWouldAddEndPuncttrue
\mciteSetBstMidEndSepPunct{\mcitedefaultmidpunct}
{\mcitedefaultendpunct}{\mcitedefaultseppunct}\relax
\EndOfBibitem
\bibitem[Csonka et~al.(2006)Csonka, Halbritter, and
  Mih\'aly]{PhysRevB.73.075405}
Csonka,~S.; Halbritter,~A.; Mih\'aly,~G. \emph{Phys. Rev. B} \textbf{2006},
  \emph{73}, 075405\relax
\mciteBstWouldAddEndPuncttrue
\mciteSetBstMidEndSepPunct{\mcitedefaultmidpunct}
{\mcitedefaultendpunct}{\mcitedefaultseppunct}\relax
\EndOfBibitem
\bibitem[Thijssen et~al.(2006)Thijssen, Marjenburgh, Bremmer, and van
  Ruitenbeek]{PhysRevLett.96.026806}
Thijssen,~W. H.~A.; Marjenburgh,~D.; Bremmer,~R.~H.; van Ruitenbeek,~J.~M.
  \emph{Phys. Rev. Lett.} \textbf{2006}, \emph{96}, 026806\relax
\mciteBstWouldAddEndPuncttrue
\mciteSetBstMidEndSepPunct{\mcitedefaultmidpunct}
{\mcitedefaultendpunct}{\mcitedefaultseppunct}\relax
\EndOfBibitem
\bibitem[Kiguchi et~al.(2007)Kiguchi, Stadler, Kristensen, Djukic, and van
  Ruitenbeek]{PhysRevLett.98.146802}
Kiguchi,~M.; Stadler,~R.; Kristensen,~I.~S.; Djukic,~D.; van Ruitenbeek,~J.~M.
  \emph{Phys. Rev. Lett.} \textbf{2007}, \emph{98}, 146802\relax
\mciteBstWouldAddEndPuncttrue
\mciteSetBstMidEndSepPunct{\mcitedefaultmidpunct}
{\mcitedefaultendpunct}{\mcitedefaultseppunct}\relax
\EndOfBibitem
\bibitem[Nakazumi and Kiguchi(2010)Nakazumi, and
  Kiguchi]{doi:10.1021/jz100084a}
Nakazumi,~T.; Kiguchi,~M. \emph{The Journal of Physical Chemistry Letters}
  \textbf{2010}, \emph{1}, 923--926\relax
\mciteBstWouldAddEndPuncttrue
\mciteSetBstMidEndSepPunct{\mcitedefaultmidpunct}
{\mcitedefaultendpunct}{\mcitedefaultseppunct}\relax
\EndOfBibitem
\bibitem[Kim et~al.(2011)Kim, Hellmuth, Bürkle, Pauly, and
  Scheer]{doi:10.1021/nn200759s}
Kim,~Y.; Hellmuth,~T.~J.; Bürkle,~M.; Pauly,~F.; Scheer,~E. \emph{ACS Nano}
  \textbf{2011}, \emph{5}, 4104--4111\relax
\mciteBstWouldAddEndPuncttrue
\mciteSetBstMidEndSepPunct{\mcitedefaultmidpunct}
{\mcitedefaultendpunct}{\mcitedefaultseppunct}\relax
\EndOfBibitem
\bibitem[Li et~al.(2008)Li, Pobelov, Wandlowski, Bagrets, Arnold, and
  Evers]{doi:10.1021/ja0762386}
Li,~C.; Pobelov,~I.; Wandlowski,~T.; Bagrets,~A.; Arnold,~A.; Evers,~F.
  \emph{J. Am. Chem. Soc.} \textbf{2008}, \emph{130}, 318--326\relax
\mciteBstWouldAddEndPuncttrue
\mciteSetBstMidEndSepPunct{\mcitedefaultmidpunct}
{\mcitedefaultendpunct}{\mcitedefaultseppunct}\relax
\EndOfBibitem
\bibitem[Venkataraman et~al.(2006)Venkataraman, Klare, Tam, Nuckolls,
  Hybertsen, and Steigerwald]{doi:10.1021/nl052373+}
Venkataraman,~L.; Klare,~J.~E.; Tam,~I.~W.; Nuckolls,~C.; Hybertsen,~M.~S.;
  Steigerwald,~M.~L. \emph{Nano Lett.} \textbf{2006}, \emph{6}, 458--462\relax
\mciteBstWouldAddEndPuncttrue
\mciteSetBstMidEndSepPunct{\mcitedefaultmidpunct}
{\mcitedefaultendpunct}{\mcitedefaultseppunct}\relax
\EndOfBibitem
\bibitem[Reed et~al.(1997)Reed, Zhou, Muller, Burgin, and Tour]{reed97}
Reed,~M.~A.; Zhou,~C.; Muller,~C.~J.; Burgin,~T.~P.; Tour,~J.~M. \emph{Science}
  \textbf{1997}, \emph{278}, 252--254\relax
\mciteBstWouldAddEndPuncttrue
\mciteSetBstMidEndSepPunct{\mcitedefaultmidpunct}
{\mcitedefaultendpunct}{\mcitedefaultseppunct}\relax
\EndOfBibitem
\bibitem[Smit et~al.(2002)Smit, Noat, Untiedt, Lang, van Hemert, and van
  Ruitenbeek]{smit02}
Smit,~R.~H.~M.; Noat,~Y.; Untiedt,~C.; Lang,~N.~D.; van Hemert,~M.~C.; van
  Ruitenbeek,~J.~M. \emph{Nature} \textbf{2002}, \emph{419}, 906--909\relax
\mciteBstWouldAddEndPuncttrue
\mciteSetBstMidEndSepPunct{\mcitedefaultmidpunct}
{\mcitedefaultendpunct}{\mcitedefaultseppunct}\relax
\EndOfBibitem
\bibitem[Xu and Tao(2003)Xu, and Tao]{xu03}
Xu,~B.; Tao,~N.~J. \emph{Science} \textbf{2003}, \emph{301}, 1221--1223\relax
\mciteBstWouldAddEndPuncttrue
\mciteSetBstMidEndSepPunct{\mcitedefaultmidpunct}
{\mcitedefaultendpunct}{\mcitedefaultseppunct}\relax
\EndOfBibitem
\bibitem[Gonz\'alez et~al.(2006)Gonz\'alez, Wu, Huber, van~der Molen,
  Schönenberger, and Calame]{NanoLett_6_2238_2006}
Gonz\'alez,~M.~T.; Wu,~S.; Huber,~R.; van~der Molen,~S.~J.; Schönenberger,~C.;
  Calame,~M. \emph{Nano Letters} \textbf{2006}, \emph{6}, 2238--2242\relax
\mciteBstWouldAddEndPuncttrue
\mciteSetBstMidEndSepPunct{\mcitedefaultmidpunct}
{\mcitedefaultendpunct}{\mcitedefaultseppunct}\relax
\EndOfBibitem
\bibitem[Mishchenko et~al.(2011)Mishchenko, Zotti, Vonlanthen, Bürkle, Pauly,
  Cuevas, Mayor, and Wandlowski]{doi:10.1021/ja107340t}
Mishchenko,~A.; Zotti,~L.~A.; Vonlanthen,~D.; Bürkle,~M.; Pauly,~F.;
  Cuevas,~J.~C.; Mayor,~M.; Wandlowski,~T. \emph{J. Am. Chem. Soc.}
  \textbf{2011}, \emph{133}, 184--187\relax
\mciteBstWouldAddEndPuncttrue
\mciteSetBstMidEndSepPunct{\mcitedefaultmidpunct}
{\mcitedefaultendpunct}{\mcitedefaultseppunct}\relax
\EndOfBibitem
\bibitem[Gonz\'{a}lez et~al.(2011)Gonz\'{a}lez, Leary, Garc\'{i}a, Verma,
  Herranz, Rubio-Bollinger, Mart\'{i}n, and Agra\"{i}t]{doi:10.1021/jp204005v}
Gonz\'{a}lez,~M.~T.; Leary,~E.; Garc\'{i}a,~R.; Verma,~P.; Herranz,~M.~n.;
  Rubio-Bollinger,~G.; Mart\'{i}n,~N.; Agra\"{i}t,~N. \emph{J. Phys. Chem. C.}
  \textbf{2011}, \emph{115}, 17973--17978\relax
\mciteBstWouldAddEndPuncttrue
\mciteSetBstMidEndSepPunct{\mcitedefaultmidpunct}
{\mcitedefaultendpunct}{\mcitedefaultseppunct}\relax
\EndOfBibitem
\bibitem[Kondo and Takayanagi(2000)Kondo, and Takayanagi]{kondo00}
Kondo,~Y.; Takayanagi,~K. \emph{Science} \textbf{2000}, \emph{289},
  606--608\relax
\mciteBstWouldAddEndPuncttrue
\mciteSetBstMidEndSepPunct{\mcitedefaultmidpunct}
{\mcitedefaultendpunct}{\mcitedefaultseppunct}\relax
\EndOfBibitem
\bibitem[Coura et~al.(2004)Coura, Legoas, Moreira, Sato, Rodrigues, Dantas,
  Ugarte, and GalvĂŁo]{doi:10.1021/nl049725h}
Coura,~P.~Z.; Legoas,~S.~B.; Moreira,~A.~S.; Sato,~F.; Rodrigues,~V.;
  Dantas,~S.~O.; Ugarte,~D.; GalvĂŁo,~D.~S. \emph{Nano Letters} \textbf{2004},
  \emph{4}, 1187--1191\relax
\mciteBstWouldAddEndPuncttrue
\mciteSetBstMidEndSepPunct{\mcitedefaultmidpunct}
{\mcitedefaultendpunct}{\mcitedefaultseppunct}\relax
\EndOfBibitem
\bibitem[Rodrigues et~al.(2000)Rodrigues, Fuhrer, and Ugarte]{rodrigues00}
Rodrigues,~V.; Fuhrer,~T.; Ugarte,~D. \emph{Phys. Rev. Lett.} \textbf{2000},
  \emph{85}, 4124--4127\relax
\mciteBstWouldAddEndPuncttrue
\mciteSetBstMidEndSepPunct{\mcitedefaultmidpunct}
{\mcitedefaultendpunct}{\mcitedefaultseppunct}\relax
\EndOfBibitem
\bibitem[Heurich et~al.(2002)Heurich, Cuevas, Wenzel, and
  Sch\"on]{PhysRevLett.88.256803}
Heurich,~J.; Cuevas,~J.~C.; Wenzel,~W.; Sch\"on,~G. \emph{Phys. Rev. Lett.}
  \textbf{2002}, \emph{88}, 256803\relax
\mciteBstWouldAddEndPuncttrue
\mciteSetBstMidEndSepPunct{\mcitedefaultmidpunct}
{\mcitedefaultendpunct}{\mcitedefaultseppunct}\relax
\EndOfBibitem
\bibitem[Jel\'\i{}nek et~al.(2006)Jel\'\i{}nek, P\'erez, Ortega, and
  Flores]{PhysRevLett.96.046803}
Jel\'\i{}nek,~P.; P\'erez,~R.; Ortega,~J.; Flores,~F. \emph{Phys. Rev. Lett.}
  \textbf{2006}, \emph{96}, 046803\relax
\mciteBstWouldAddEndPuncttrue
\mciteSetBstMidEndSepPunct{\mcitedefaultmidpunct}
{\mcitedefaultendpunct}{\mcitedefaultseppunct}\relax
\EndOfBibitem
\bibitem[Mishchenko et~al.(2010)Mishchenko, Vonlanthen, Meded, BuĚrkle, Li,
  Pobelov, Bagrets, Viljas, Pauly, Evers, Mayor, and
  Wandlowski]{Mischenko_Nanolett_10_156_2010}
Mishchenko,~A.; Vonlanthen,~D.; Meded,~V.; BuĚrkle,~M.; Li,~C.;
  Pobelov,~I.~V.; Bagrets,~A.; Viljas,~J.~K.; Pauly,~F.; Evers,~F.; Mayor,~M.;
  Wandlowski,~T. \emph{Nano Letters} \textbf{2010}, \emph{10}, 156--163\relax
\mciteBstWouldAddEndPuncttrue
\mciteSetBstMidEndSepPunct{\mcitedefaultmidpunct}
{\mcitedefaultendpunct}{\mcitedefaultseppunct}\relax
\EndOfBibitem
\bibitem[Martin et~al.(2010)Martin, Grace, Bryce, Wang, Jitchati, Batsanov,
  Higgins, Lambert, and Nichols]{doi:10.1021/ja103327f}
Martin,~S.; Grace,~I.; Bryce,~M.~R.; Wang,~C.; Jitchati,~R.; Batsanov,~A.~S.;
  Higgins,~S.~J.; Lambert,~C.~J.; Nichols,~R.~J. \emph{Journal of the American
  Chemical Society} \textbf{2010}, \emph{132}, 9157--9164\relax
\mciteBstWouldAddEndPuncttrue
\mciteSetBstMidEndSepPunct{\mcitedefaultmidpunct}
{\mcitedefaultendpunct}{\mcitedefaultseppunct}\relax
\EndOfBibitem
\bibitem[Makk et~al.(2011)Makk, Visontai, Oroszl\'any, Manrique, Csonka,
  Cserti, C., and Halbritter]{indium}
Makk,~P.; Visontai,~D.; Oroszl\'any,~L.; Manrique,~D.~Z.; Csonka,~S.;
  Cserti,~J.; C.,~L.; Halbritter,~A. \emph{Phys. Rev. Lett.} \textbf{2011},
  \emph{107}, 276801\relax
\mciteBstWouldAddEndPuncttrue
\mciteSetBstMidEndSepPunct{\mcitedefaultmidpunct}
{\mcitedefaultendpunct}{\mcitedefaultseppunct}\relax
\EndOfBibitem
\bibitem[van~den Brom and van Ruitenbeek(1999)van~den Brom, and van
  Ruitenbeek]{PhysRevLett.82.1526}
van~den Brom,~H.~E.; van Ruitenbeek,~J.~M. \emph{Phys. Rev. Lett.}
  \textbf{1999}, \emph{82}, 1526--1529\relax
\mciteBstWouldAddEndPuncttrue
\mciteSetBstMidEndSepPunct{\mcitedefaultmidpunct}
{\mcitedefaultendpunct}{\mcitedefaultseppunct}\relax
\EndOfBibitem
\bibitem[Djukic and van Ruitenbeek(2006)Djukic, and van
  Ruitenbeek]{doi:10.1021/nl060116e}
Djukic,~D.; van Ruitenbeek,~J.~M. \emph{Nano Letters} \textbf{2006}, \emph{6},
  789--793\relax
\mciteBstWouldAddEndPuncttrue
\mciteSetBstMidEndSepPunct{\mcitedefaultmidpunct}
{\mcitedefaultendpunct}{\mcitedefaultseppunct}\relax
\EndOfBibitem
\bibitem[Wheeler et~al.(2010)Wheeler, Russom, Evans, King, and
  Natelson]{doi:10.1021/nl904052r}
Wheeler,~P.~J.; Russom,~J.~N.; Evans,~K.; King,~N.~S.; Natelson,~D. \emph{Nano
  Letters} \textbf{2010}, \emph{10}, 1287--1292, PMID: 20205414\relax
\mciteBstWouldAddEndPuncttrue
\mciteSetBstMidEndSepPunct{\mcitedefaultmidpunct}
{\mcitedefaultendpunct}{\mcitedefaultseppunct}\relax
\EndOfBibitem
\bibitem[Ludoph et~al.(1999)Ludoph, Devoret, Esteve, Urbina, and van
  Ruitenbeek]{ludoph99}
Ludoph,~B.; Devoret,~M.~H.; Esteve,~D.; Urbina,~C.; van Ruitenbeek,~J.~M.
  \emph{Phys. Rev. Lett.} \textbf{1999}, \emph{82}, 1530--1533\relax
\mciteBstWouldAddEndPuncttrue
\mciteSetBstMidEndSepPunct{\mcitedefaultmidpunct}
{\mcitedefaultendpunct}{\mcitedefaultseppunct}\relax
\EndOfBibitem
\bibitem[Halbritter et~al.(2004)Halbritter, Csonka, Mih\'aly, Shklyarevskii,
  Speller, and van Kempen]{PhysRevB.69.121411}
Halbritter,~A.; Csonka,~S.; Mih\'aly,~G.; Shklyarevskii,~O.~I.; Speller,~S.;
  van Kempen,~H. \emph{Phys. Rev. B} \textbf{2004}, \emph{69}, 121411\relax
\mciteBstWouldAddEndPuncttrue
\mciteSetBstMidEndSepPunct{\mcitedefaultmidpunct}
{\mcitedefaultendpunct}{\mcitedefaultseppunct}\relax
\EndOfBibitem
\bibitem[Scheer et~al.(1998)Scheer, {Agra\"{\i}t}, Cuevas, {Levy Yeyati},
  Ludoph, Mart{\'\i}n-Rodero, {Rubio Bollinger}, van Ruitenbeek, and
  Urbina]{scheer98}
Scheer,~E.; {Agra\"{\i}t},~N.; Cuevas,~J.~C.; {Levy Yeyati},~A.; Ludoph,~B.;
  Mart{\'\i}n-Rodero,~A.; {Rubio Bollinger},~G.; van Ruitenbeek,~J.~M.;
  Urbina,~C. \emph{Nature} \textbf{1998}, \emph{394}, 154--157\relax
\mciteBstWouldAddEndPuncttrue
\mciteSetBstMidEndSepPunct{\mcitedefaultmidpunct}
{\mcitedefaultendpunct}{\mcitedefaultseppunct}\relax
\EndOfBibitem
\bibitem[Makk et~al.(2008)Makk, Csonka, and Halbritter]{makk08}
Makk,~P.; Csonka,~S.; Halbritter,~A. \emph{Phys. Rev. B} \textbf{2008},
  \emph{78}, 045414\relax
\mciteBstWouldAddEndPuncttrue
\mciteSetBstMidEndSepPunct{\mcitedefaultmidpunct}
{\mcitedefaultendpunct}{\mcitedefaultseppunct}\relax
\EndOfBibitem
\bibitem[Ludoph and van Ruitenbeek(1999)Ludoph, and van Ruitenbeek]{ludoph99a}
Ludoph,~B.; van Ruitenbeek,~J.~M. \emph{Phys. Rev. B} \textbf{1999}, \emph{59},
  12290--12293\relax
\mciteBstWouldAddEndPuncttrue
\mciteSetBstMidEndSepPunct{\mcitedefaultmidpunct}
{\mcitedefaultendpunct}{\mcitedefaultseppunct}\relax
\EndOfBibitem
\bibitem[Koch et~al.(2004)Koch, von Oppen, Oreg, and Sela]{PhysRevB.70.195107}
Koch,~J.; von Oppen,~F.; Oreg,~Y.; Sela,~E. \emph{Phys. Rev. B} \textbf{2004},
  \emph{70}, 195107\relax
\mciteBstWouldAddEndPuncttrue
\mciteSetBstMidEndSepPunct{\mcitedefaultmidpunct}
{\mcitedefaultendpunct}{\mcitedefaultseppunct}\relax
\EndOfBibitem
\bibitem[Tal et~al.(2008)Tal, Krieger, Leerink, and van
  Ruitenbeek]{PhysRevLett.100.196804}
Tal,~O.; Krieger,~M.; Leerink,~B.; van Ruitenbeek,~J.~M. \emph{Phys. Rev.
  Lett.} \textbf{2008}, \emph{100}, 196804\relax
\mciteBstWouldAddEndPuncttrue
\mciteSetBstMidEndSepPunct{\mcitedefaultmidpunct}
{\mcitedefaultendpunct}{\mcitedefaultseppunct}\relax
\EndOfBibitem
\bibitem[Arroyo et~al.(2010)Arroyo, Frederiksen, Rubio-Bollinger, V\'elez,
  Arnau, S\'anchez-Portal, and Agra\"\i{}t]{PhysRevB.81.075405}
Arroyo,~C.~R.; Frederiksen,~T.; Rubio-Bollinger,~G.; V\'elez,~M.; Arnau,~A.;
  S\'anchez-Portal,~D.; Agra\"\i{}t,~N. \emph{Phys. Rev. B} \textbf{2010},
  \emph{81}, 075405\relax
\mciteBstWouldAddEndPuncttrue
\mciteSetBstMidEndSepPunct{\mcitedefaultmidpunct}
{\mcitedefaultendpunct}{\mcitedefaultseppunct}\relax
\EndOfBibitem
\bibitem[Martin et~al.(2008)Martin, Ding, S\"orensen, Bj\"ornholm, van
  Ruitenbeek, and van~der Zant]{Martin_JACS_130_13198_2008}
Martin,~C.~A.; Ding,~D.; S\"orensen,~J.~K.; Bj\"ornholm,~T.; van
  Ruitenbeek,~J.~M.; van~der Zant,~H. S.~J. \emph{Journal of the American
  Chemical Society} \textbf{2008}, \emph{130}, 13198--13199\relax
\mciteBstWouldAddEndPuncttrue
\mciteSetBstMidEndSepPunct{\mcitedefaultmidpunct}
{\mcitedefaultendpunct}{\mcitedefaultseppunct}\relax
\EndOfBibitem
\bibitem[Quek et~al.(2009)Quek, Kamenetska, Steigerwald, Choi, Louie,
  Hybertsen, Neaton, and Venkataraman]{Quek2009}
Quek,~S.~Y.; Kamenetska,~M.; Steigerwald,~M.~L.; Choi,~H.~J.; Louie,~S.~G.;
  Hybertsen,~M.~S.; Neaton,~J.~B.; Venkataraman,~L. \emph{Nat Nano}
  \textbf{2009}, \emph{4}, 230--234\relax
\mciteBstWouldAddEndPuncttrue
\mciteSetBstMidEndSepPunct{\mcitedefaultmidpunct}
{\mcitedefaultendpunct}{\mcitedefaultseppunct}\relax
\EndOfBibitem
\bibitem[Halbritter et~al.(2010)Halbritter, Makk, Mackowiak, Csonka,
  Wawrzyniak, and Martinek]{PhysRevLett.105.266805}
Halbritter,~A.; Makk,~P.; Mackowiak,~S.; Csonka,~S.; Wawrzyniak,~M.;
  Martinek,~J. \emph{Phys. Rev. Lett.} \textbf{2010}, \emph{105}, 266805\relax
\mciteBstWouldAddEndPuncttrue
\mciteSetBstMidEndSepPunct{\mcitedefaultmidpunct}
{\mcitedefaultendpunct}{\mcitedefaultseppunct}\relax
\EndOfBibitem
\bibitem[Makk et~al.(2012)Makk, Tomaszewski, Martinek, Balogh, Csonka,
  Wawrzyniak, Frei, Venkataraman, and Halbritter]{doi:10.1021/nn300440f}
For a detailed review see: Makk,~P.; Tomaszewski,~D.; Martinek,~J.; Balogh,~Z.; Csonka,~S.;
  Wawrzyniak,~M.; Frei,~M.; Venkataraman,~L.; Halbritter,~A. \emph{ACS Nano}
  \textbf{2012}, \emph{6}, 3411--3423\relax
\mciteBstWouldAddEndPuncttrue
\mciteSetBstMidEndSepPunct{\mcitedefaultmidpunct}
{\mcitedefaultendpunct}{\mcitedefaultseppunct}\relax
\EndOfBibitem
\bibitem[Tal et~al.(2009)Tal, Kiguchi, Thijssen, Djukic, Untiedt, Smit, and van
  Ruitenbeek]{PhysRevB.80.085427}
Tal,~O.; Kiguchi,~M.; Thijssen,~W. H.~A.; Djukic,~D.; Untiedt,~C.; Smit,~R.
  H.~M.; van Ruitenbeek,~J.~M. \emph{Phys. Rev. B} \textbf{2009}, \emph{80},
  085427\relax
\mciteBstWouldAddEndPuncttrue
\mciteSetBstMidEndSepPunct{\mcitedefaultmidpunct}
{\mcitedefaultendpunct}{\mcitedefaultseppunct}\relax
\EndOfBibitem
\bibitem[Kiguchi et~al.(2007)Kiguchi, Djukic, and van
  Ruitenbeek]{0957-4484-18-3-035205}
Kiguchi,~M.; Djukic,~D.; van Ruitenbeek,~J.~M. \emph{Nanotechnology}
  \textbf{2007}, \emph{18}, 035205\relax
\mciteBstWouldAddEndPuncttrue
\mciteSetBstMidEndSepPunct{\mcitedefaultmidpunct}
{\mcitedefaultendpunct}{\mcitedefaultseppunct}\relax
\EndOfBibitem
\bibitem[Strange et~al.(2006)Strange, Thygesen, and
  Jacobsen]{PhysRevB.73.125424}
Strange,~M.; Thygesen,~K.~S.; Jacobsen,~K.~W. \emph{Phys. Rev. B}
  \textbf{2006}, \emph{73}, 125424\relax
\mciteBstWouldAddEndPuncttrue
\mciteSetBstMidEndSepPunct{\mcitedefaultmidpunct}
{\mcitedefaultendpunct}{\mcitedefaultseppunct}\relax
\EndOfBibitem
\bibitem[Nielsen et~al.(2003)Nielsen, Noat, Brandbyge, Smit, Hansen, Chen,
  Yanson, Besenbacher, and van Ruitenbeek]{nielsen03}
Nielsen,~S.~K.; Noat,~Y.; Brandbyge,~M.; Smit,~R.~H.~M.; Hansen,~K.;
  Chen,~L.~Y.; Yanson,~A.~I.; Besenbacher,~F.; van Ruitenbeek,~J.~M.
  \emph{Phys. Rev. B.} \textbf{2003}, \emph{67}, 245411\relax
\mciteBstWouldAddEndPuncttrue
\mciteSetBstMidEndSepPunct{\mcitedefaultmidpunct}
{\mcitedefaultendpunct}{\mcitedefaultseppunct}\relax
\EndOfBibitem
\bibitem[not()]{note}
The long $0.5\,$G$_0$ plateau in the 2DCDH may as well results from shorter
  $0.5\,$G$_0$ plateaus starting and breaking at various electrode
  displacements. To exlude this artifact we have also constructed a 2DCDH in
  which the displacement origin of the traces is set to the breaking point,
  ($G_{\mathrm{ref}}=0.1\,$G$_0$), and which is constructed only from the
  traces breaking from the ($0.5\,$G$_0$) configuration. This analysis
  justifies that the parallel molecular configuration can indeed pull $5\,$\AA~
  long chains.\relax
\mciteBstWouldAddEndPunctfalse
\mciteSetBstMidEndSepPunct{\mcitedefaultmidpunct}
{}{\mcitedefaultseppunct}\relax
\EndOfBibitem
\end{mcitethebibliography}
\end{document}